\documentclass[12pt]{article}
\usepackage{amsmath,amssymb}
\usepackage[includeheadfoot,margin=1in]{geometry}
\usepackage{amsmath,amssymb,units}
\usepackage[dvipdfmx]{graphicx}
\usepackage[export]{adjustbox}
\usepackage{comment}
\usepackage{listings}
\usepackage[font=small,labelfont=bf,tableposition=top]{caption}
\usepackage{bm}
\usepackage{authblk}
\usepackage{hyperref}
\usepackage{gensymb}
\usepackage{physics}
\numberwithin{equation}{section}
\newcommand{\un}[1]{\,\unit{#1}}
\begin{document}

\title{A novel water-Cherenkov detector design with retro-reflectors to produce antipodal rings\footnote{Typeset proceedings of the poster presented at the XXVIII International Conference on Neutrino Physics and Astrophysics (Neutrino 2018). The poster is available at \url{http://doi.org/10.5281/zenodo.1288376}}}
\author{Lukas Berns}
\affil{Department of Physics, Tokyo Institute of Technology\\2-12-1 Ookayama, Meguro-ku, Tokyo 152-8550, Japan}
\renewcommand\Affilfont{\itshape\small}
\date{}
\maketitle

\abstract{
Since Kamiokande, the basic design of water-Cherenkov detectors has not changed: the walls of a water tank are lined with photodetectors that capture Cherenkov photons produced by relativistic particles. However, with this design the majority of photons are lost in insensitive regions between photodetectors, while at the same time most photodetectors are outside the ring and remain dark. To fix both issues at once, we propose fixing retro-reflectors between all photodetectors. These devices will reflect uncollected photons back through their emission point onto photodetectors at the other side of the tank, producing a secondary, delayed Cherenkov ring. Numerical simulations show that, due to the parallax effect of this antipodal ring, our system can yield up to 2x improvement of detector vertex and angle resolutions. This improvement would be beneficial for kinematic selection of multi-ring events and would lower detector costs by decreasing the number of required photodetectors.
}

\pagebreak

\section{Background}
\label{sec:background}

\begin{figure}
\begin{center}
\includegraphics[valign=c,width=1.95in]{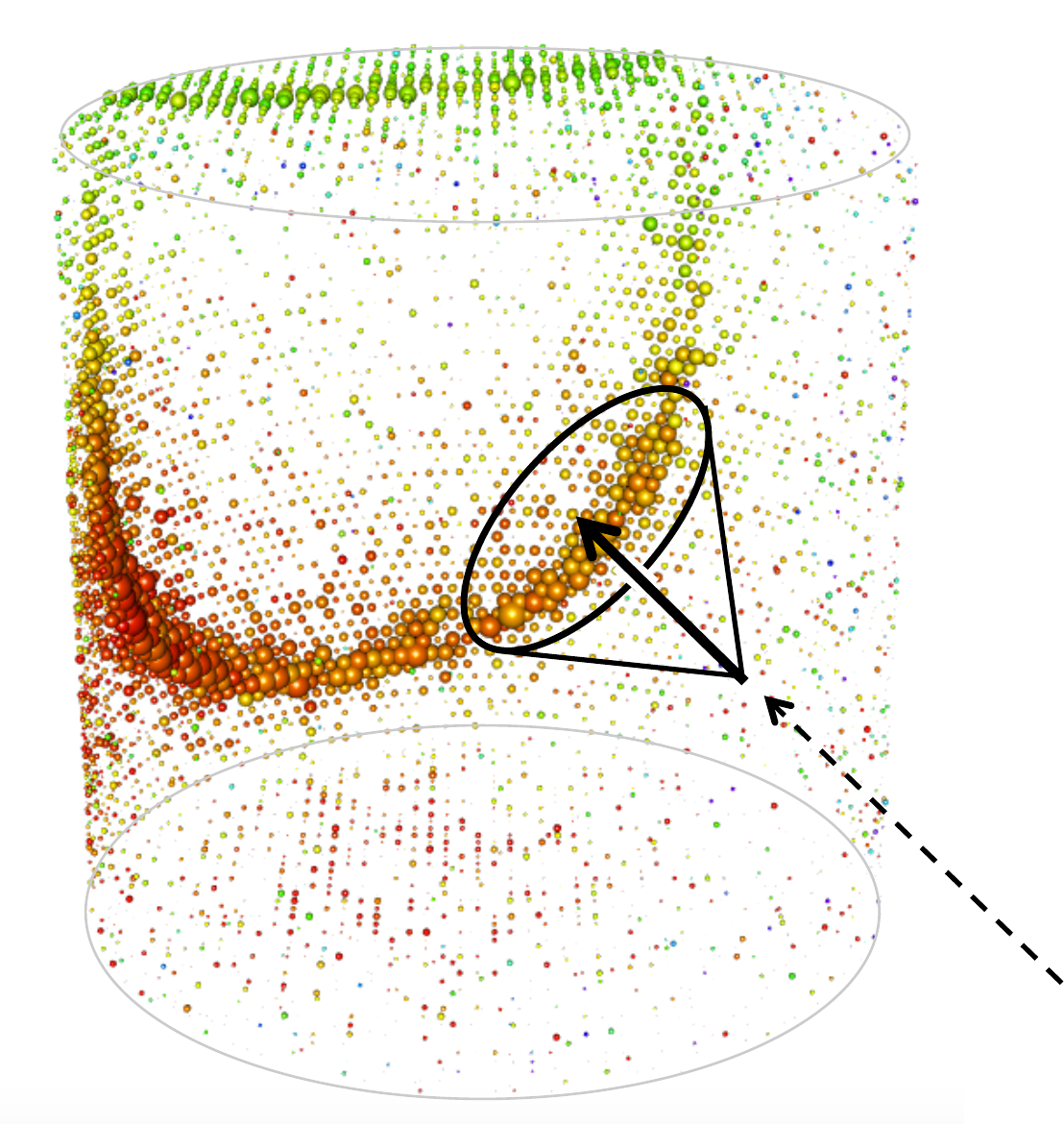}
\includegraphics[valign=c,width=1.8in]{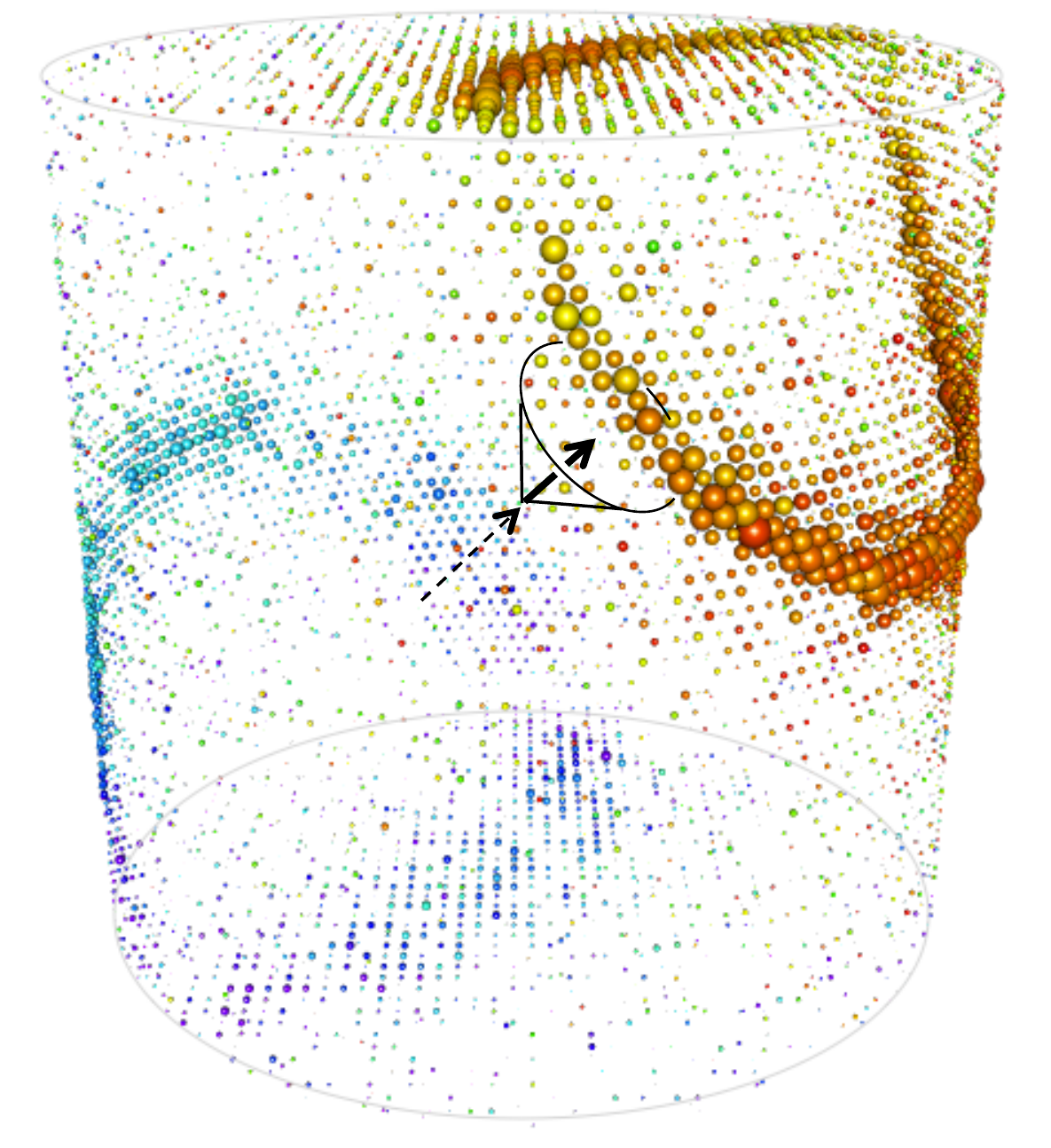}
\caption{Visualization of typical events in the conventional design (left) and in the proposed detector design using retro-reflectors (right), simulated using the analysis method described in Sec.~\ref{sec:sensitivity}. The spheres represent the PMT response with the size being related to the collected charge, and color to the trigger time. With retro-reflectors one can clearly see an antipodal ring well separated in time.}
\label{fig:01-cherenkov}
\end{center}
\end{figure}

The basic principle of water-Cherenkov detectors is well established, and has not changed much since e.g. Kamiokande~\cite{Kamiokande}, through Super-Kamiokande~\cite{SK} (SuperK), to the currently planned Hyper-Kamiokande~\cite{HKDR} experiments (a notable exception being the addition of Gadolinium which is being prepared right now). One wonders if there is still room for improvement. What stands out are the 60\% of photons that are being lost due to the limited photo-coverage of about 40\%. At the same time, a typical ring event at $500 \un{MeV}$ (roughly the typical energy for the T2K experiment~\cite{T2K}) registers hits to about 10\% of all photomultiplier-tubes (PMTs), the remaining 90\% being ``unused" for the ring reconstruction. It is tempting to think about the possibility of mapping the photons lost by falling between the PMTs, onto the PMTs on the other side of the tank.

A naive solution would be to replace the blacksheets between the PMTs with mirrors. This however creates serious problems with reconstruction, since with a typical absorption length of $100\un{m}$ inside water, we would need to simulate about 2--3 reflections, which is computationally very expensive. One also needs to precisely align the mirrors, since even slight misalignments of order $1^\circ$ can cause order $1\un{m}$ deviations of the light path on the other side of the tank. Furthermore, it is hard to separate the reflected light based on timing alone, so the reflections would most certainly cause issues in the reconstruction of multi-ring events.

\begin{figure}
\begin{center}
\begin{small}
a.~\includegraphics[width=2.5in]{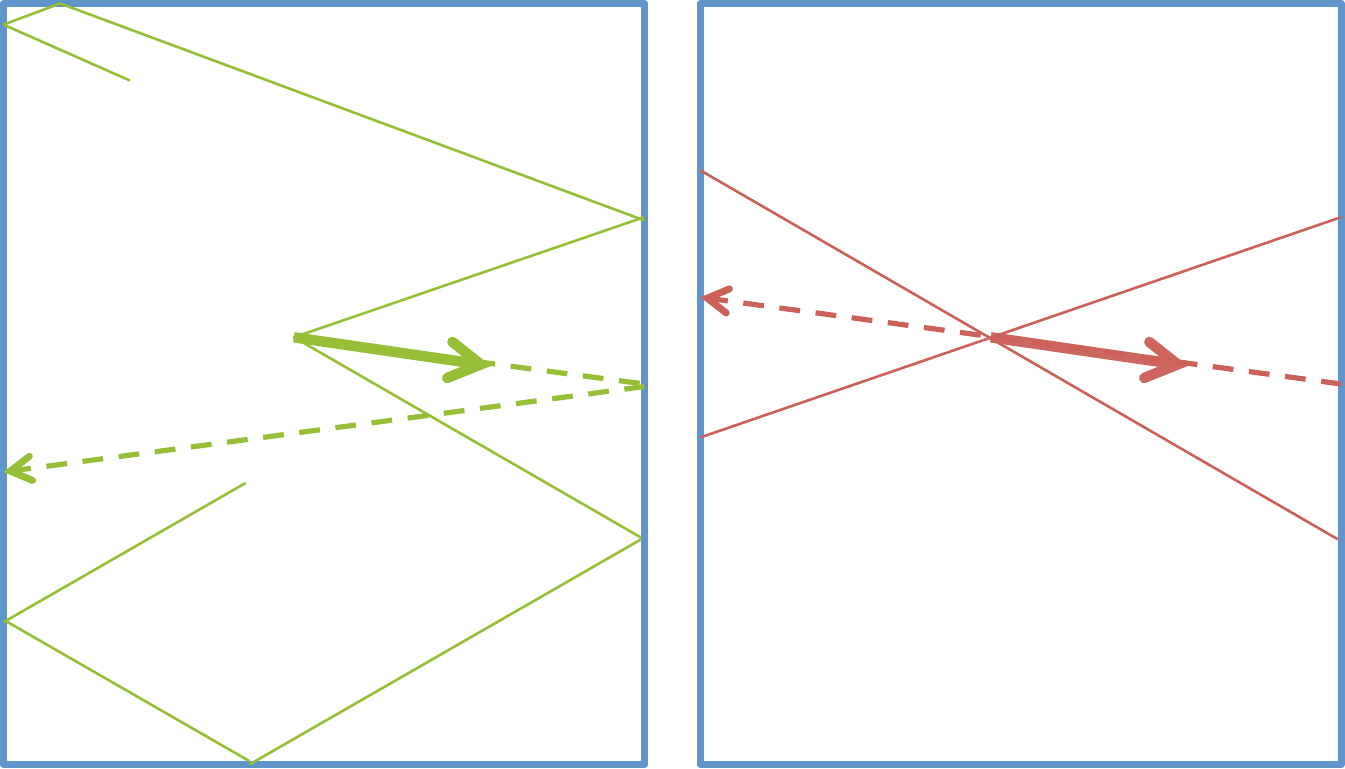} \hspace{1cm}
b.~\includegraphics[width=2.5in]{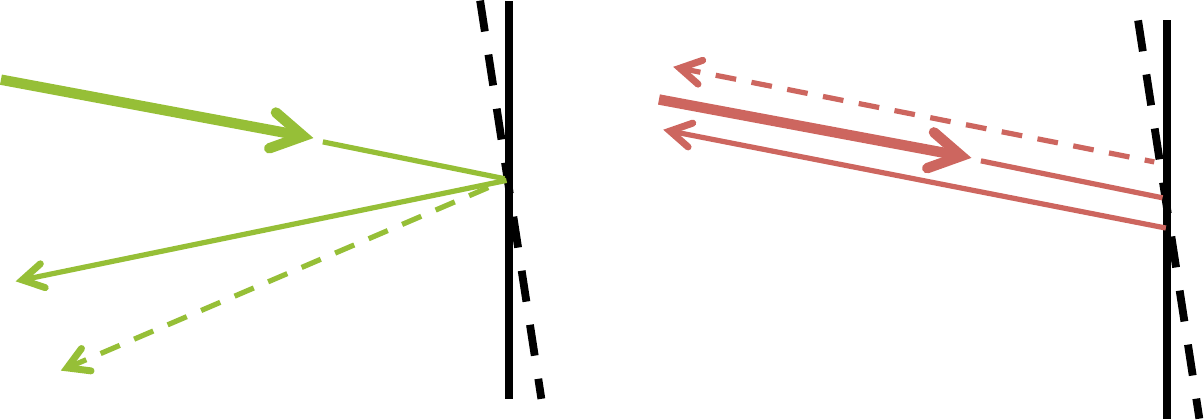}
\end{small}
\caption{a.~Reflected light path for normal mirrors (left) and retro-reflectors (right). b.~Angle of reflected light when slightly changing the alignment.}
\label{fig:02-mirror-problems}
\end{center}
\end{figure}

\section{Antipodal rings from retro-reflectors}
\label{sec:antipodal}

We propose the use of retro-reflectors to circumvent these issues. Retro-reflectors such as corner cubes reflect the light back into the incident direction. Thus the Cherenkov light emitted from the tracks would get reflected back through the emission point onto the other side of the tank, creating a second, anti-podal ring (Fig.~\ref{fig:01-cherenkov}). Even if the light hits a retro-reflector again, ideally it would get trapped between the two retro-reflectors, never to be seen again. Thus only a single reflection needs to be considered. Since the retro-reflection does not depend on the reflector alignment, this method is robust to misalignments apart from slight changes in the ``acceptance" of the reflectors.

\section{Validation of multi-reflection cancellation}

\begin{figure}
\begin{center}
\begin{small}
\includegraphics[width=6.5in]{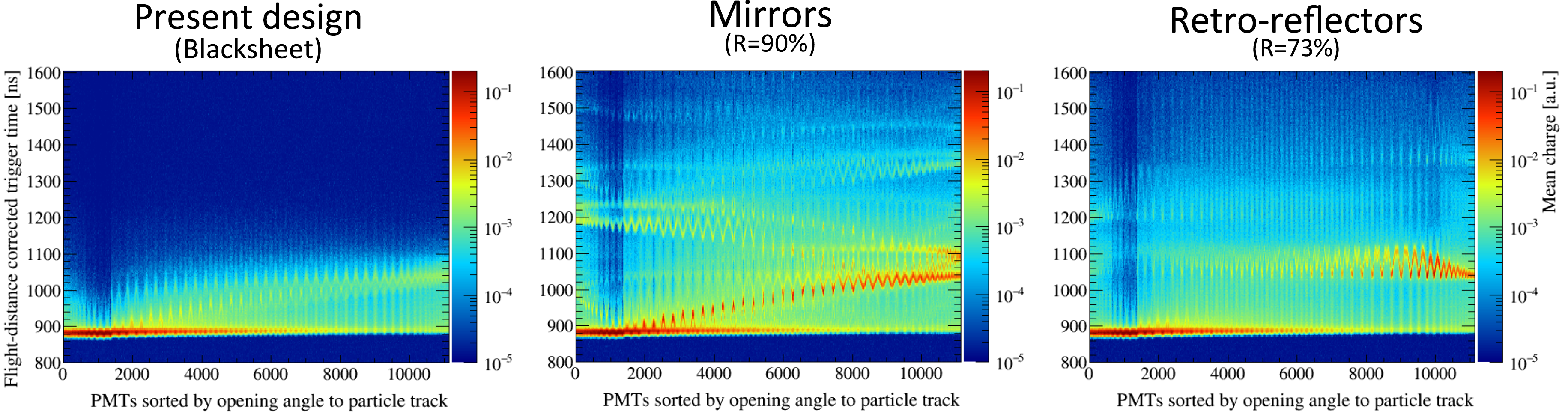}
\end{small}
\caption{Photon hit timing distribution for different detector designs.
The time has been corrected for the flight distance (distance from emission point to PMT), so direct light appears as a horizontal line. The PMTs are sorted by opening angle to the particle track, where after binning the opening angle in 30 bins, the PMTs are further sorted by the polar angle around the particle track, spiraling from the tip of the particle track to the back. In linear scale (not shown here) one would be able to identify the Cherenkov peak around PMT 1500. Since the timing of reflected light contains geometrical effects that cannot be corrected by subtracting the flight distance from the emission point, they appear as sinusoidal lines in this plot.
}
\label{fig:03-timing-distribution}
\end{center}
\end{figure}

Since Rayleigh scattering in water and ordinary reflection on PMT surfaces can break the idealized trapping of multi-reflected light between two retro-reflectors, we performed a test using the Geant4\cite{Geant4}-based simulator WCSim~\cite{WCSim}. The blacksheets normally placed between the PMTs are replaced by ordinary mirrors (90\% reflectivity) or ideal retro-reflectors (73\% reflectivity). Simulating 20,000 events of a single side-going electron of $500\un{MeV}$, we plot the trigger timings of the PMTs weighted by their charge for each PMT (Fig.~\ref{fig:03-timing-distribution}). In the conventional configuration using blacksheets, the timings are dominated by the initial photon hits (horizontal line) with a faint reflection toward the back (diagonal sinusoid). The introduction of ordinary mirrors causes very complicated reflection patterns, with light traveling back and forth a couple of times. With retro-reflectors however, we have after the direct photon ring (horizontal line), a secondary ring delayed by about 150--200~ns (sinusoid), and that's mostly the full story. Only a faint contribution from photons reflecting twice or three times appears as blurry horizontal lines approximately equally separated in time. We also note that the antipodal ring is well separated in time, such that we do not need to worry about losing reconstruction performance for the direct light ring. So (at least for ideal) retro-reflectors, the reflections are well controlled and it should not be hard to model these for reconstruction.

\section{Sensitivity estimation}
\label{sec:sensitivity}


Using the Cherenkov profile model defined in the next paragraph, the expected charge and timing distributions for all PMTs are calculated for a set of track parameters
$\eta = (t, x, y, z, E, \theta, \phi)$ and detector configuration. This is used to construct a likelihood function $L(\eta,\eta_\text{true})$, whose second derivative is used to compute the ring reconstruction sensitivities. The tank is assumed to have the size of the SuperK inner detector with 40\% photo-coverage.


\begin{figure}
\begin{center}
\begin{small}
\includegraphics[width=5in]{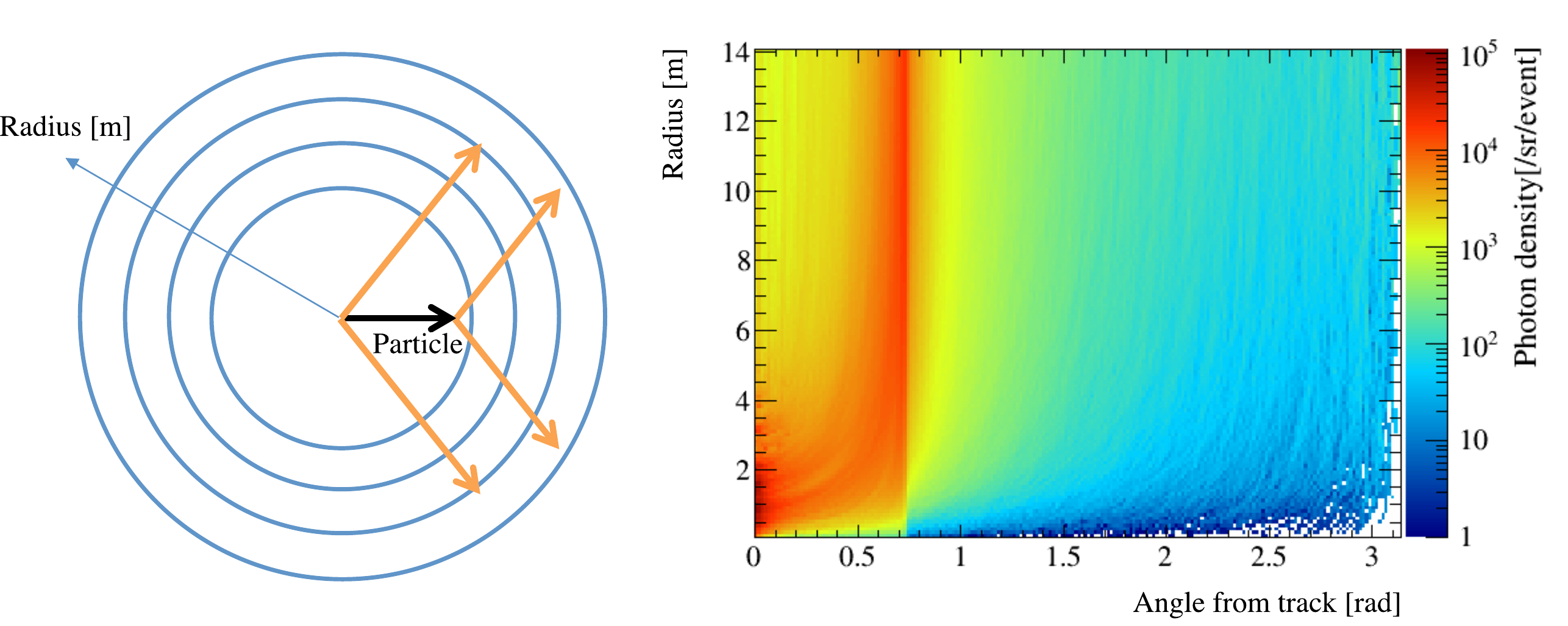}
\end{small}
\caption{Left: illustration of the radial bins at which the photon distribution is measured. Right: photon flux of a 500 MeV electron simulated with Geant4. At distances less than about $3\un{m}$ the finite track length means photons can be found at $0\un{rad}$. With increasing distance the radial dependence disappears since the track becomes more like a point-source, with the Cherenkov distribution showing the characteristic peak at $\cos \theta = 1/n\beta$.}
\label{fig:04-cherenkov-profile}
\end{center}
\end{figure}

For precise modeling of the Cherenkov profile, we simulate moving particles in Geant4, and fit flux and timing distributions of photons with zonal harmonics for each radial slice about the vertex (Fig.~\ref{fig:04-cherenkov-profile}). The time distribution at each solid angle is parametrized by a gaussian distribution, while for the photon flux we also model the directional distribution using a Kent distribution~\cite{Kent}, to make sure reflection and scattering effects can be implemented correctly. It is also easily possible to introduce gaussian spread due to imperfections of the retro-reflectors by modifying the parameters of the Kent distribution. The energy dependence is obtained by repeating this process for different incident energies and interpolating. For the following analysis however only 500 MeV electrons were used, so a simple linear dependence of the photon flux on the kinetic energy was assumed.

\begin{figure}
\begin{center}
\begin{small}
\includegraphics[valign=t,width=3in]{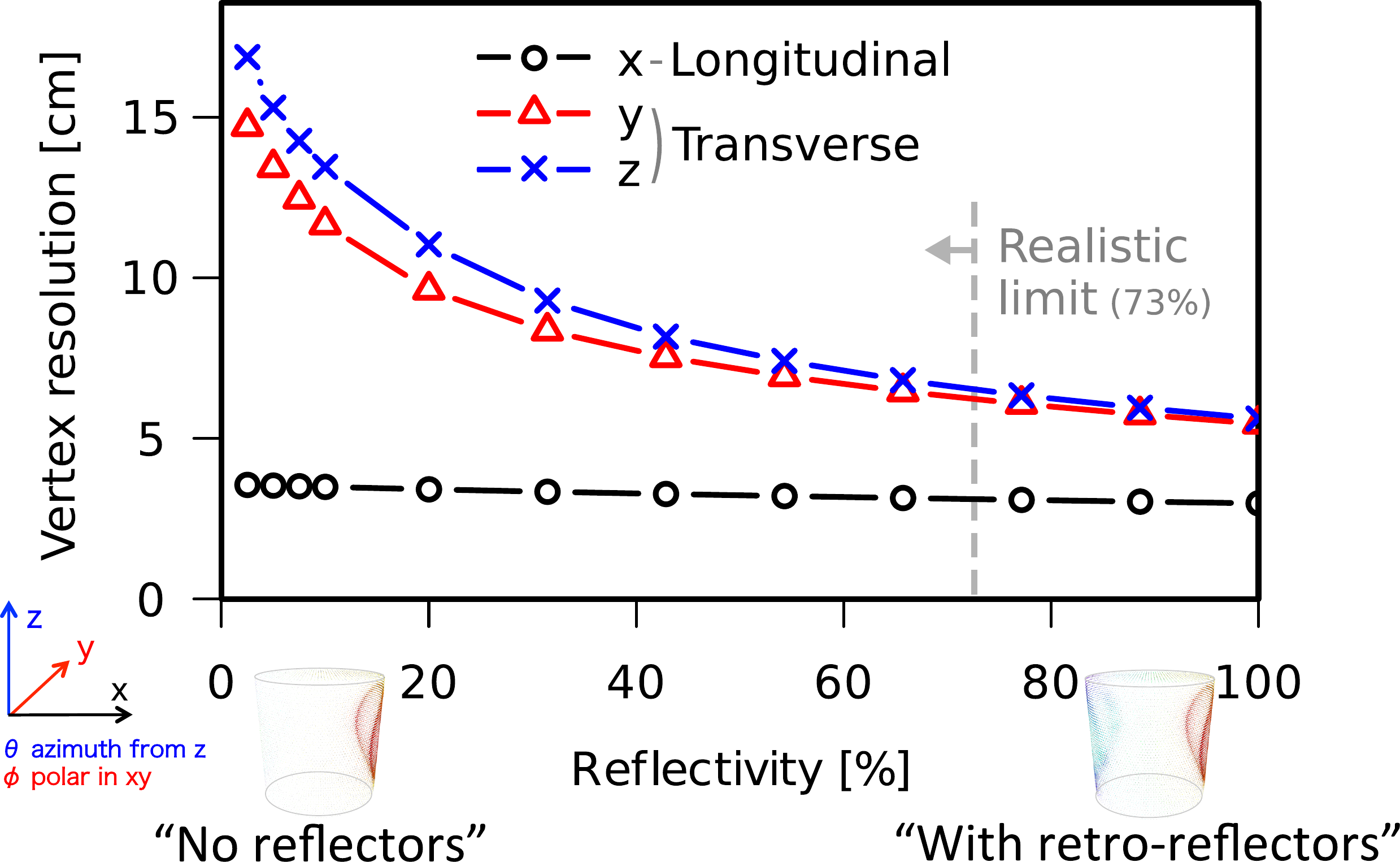}
\includegraphics[valign=t,width=2.6in]{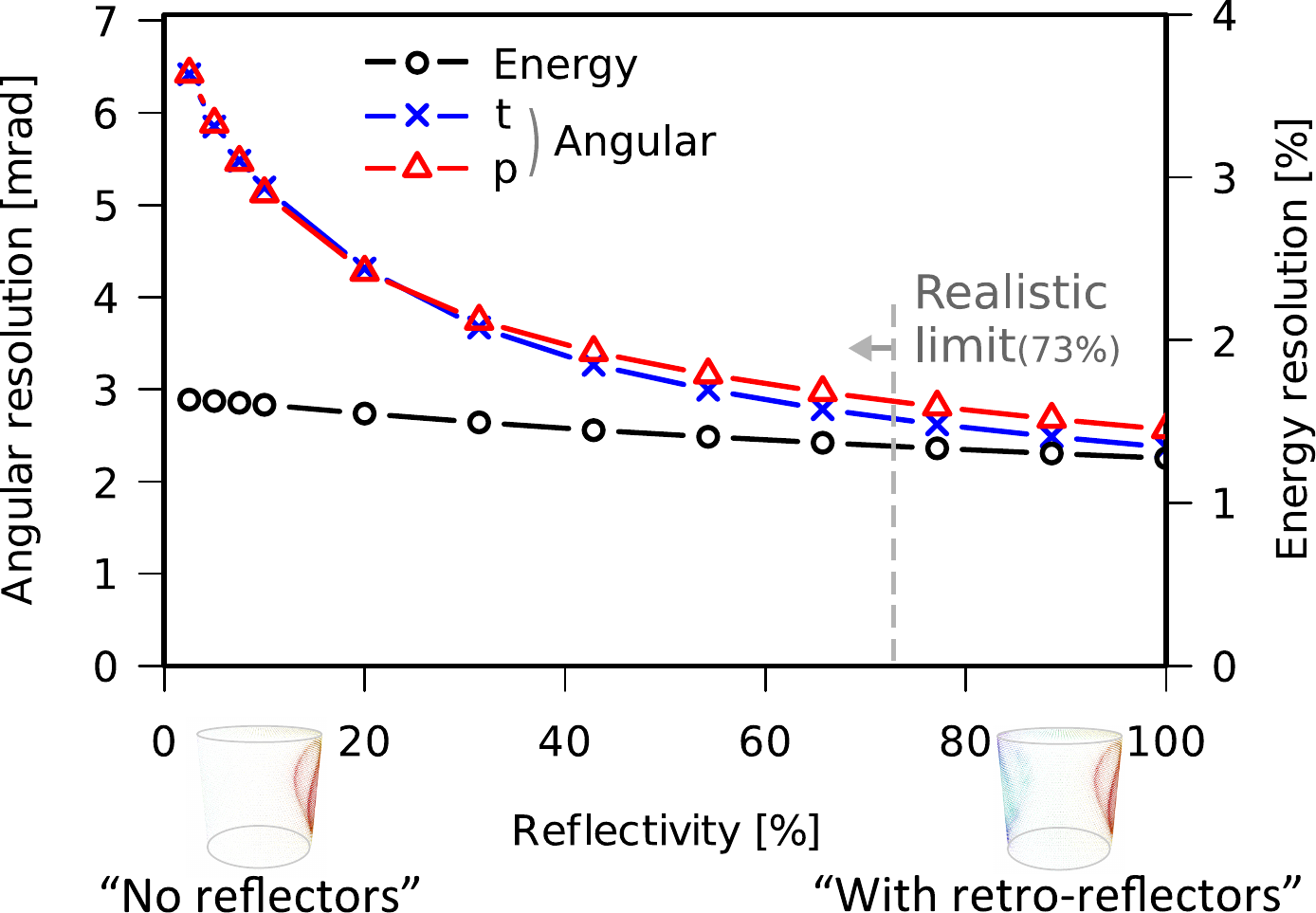}
\end{small}
\caption{Vertex, energy, and angular resolution for a 500 MeV electron emitted from the center of the tank toward the side, calculated for different values of reflectivity for the reflectors.}
\label{fig:05-vertex-resolution}
\end{center}
\end{figure}

The calculated vertex, energy, and angular resolutions for varying reflectivity are shown in Fig.~\ref{fig:05-vertex-resolution}. Since even in SuperK one has reflections from PMT surfaces and blacksheets, we expect a reflectivity of a few percent to correspond to realistic SuperK performance. For corner cube retro-reflectors a realistic reflectivity is below 73\% (triple reflection on 95\% reflectivity Aluminum mirrors). We see that for increasing reflectivity the vertex resolution in the transverse direction ($y$, $z$ --- the electron is taken to move into the $x$ direction) improves significantly up to about factor 2. The directional resolution (azimuth $\theta$ and polar $\phi$ about the $z$ axis) shows a very similar improvement which as we discuss in the next section is not a coincidence. The longitudinal vertex ($x$) and energy ($E$) resolution only improve slightly.

\section{Origin of enhanced resolution}

\begin{figure}
\begin{center}
\begin{small}
a.~\includegraphics[height=1.5in]{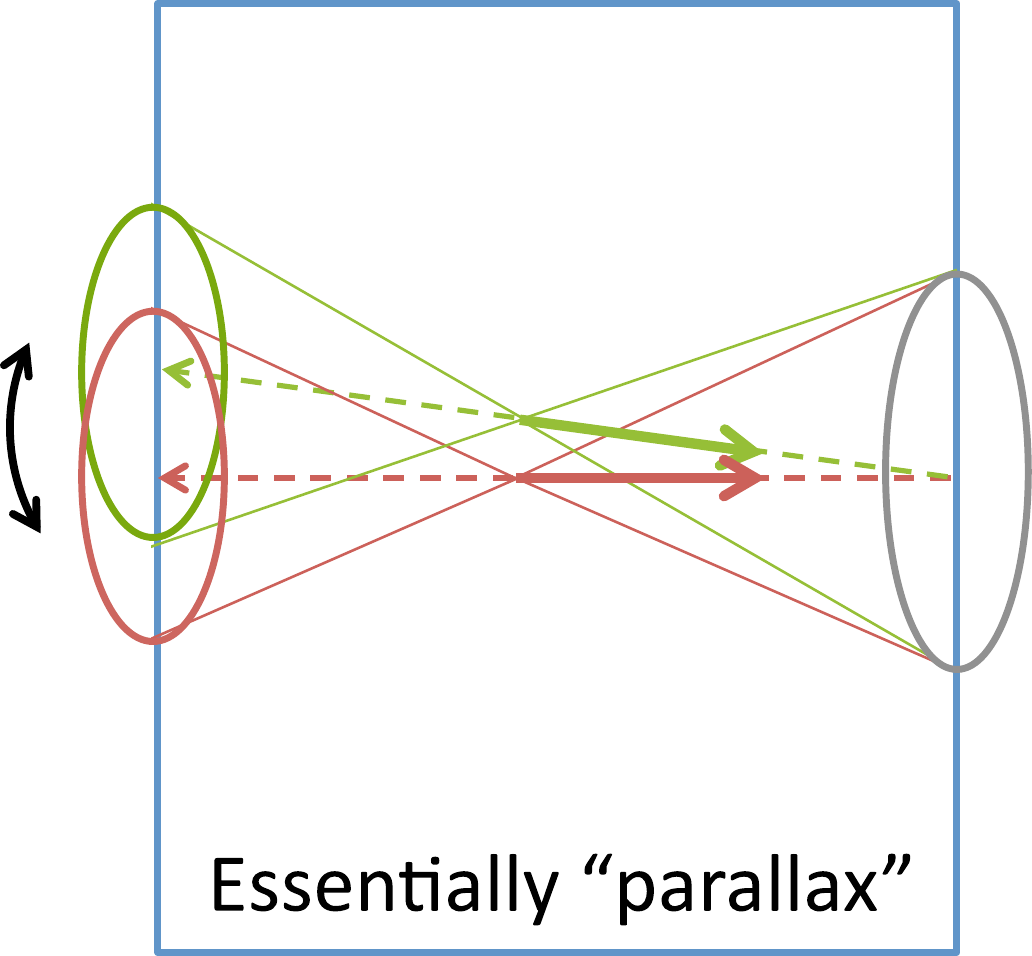}
b.~\includegraphics[height=1.5in]{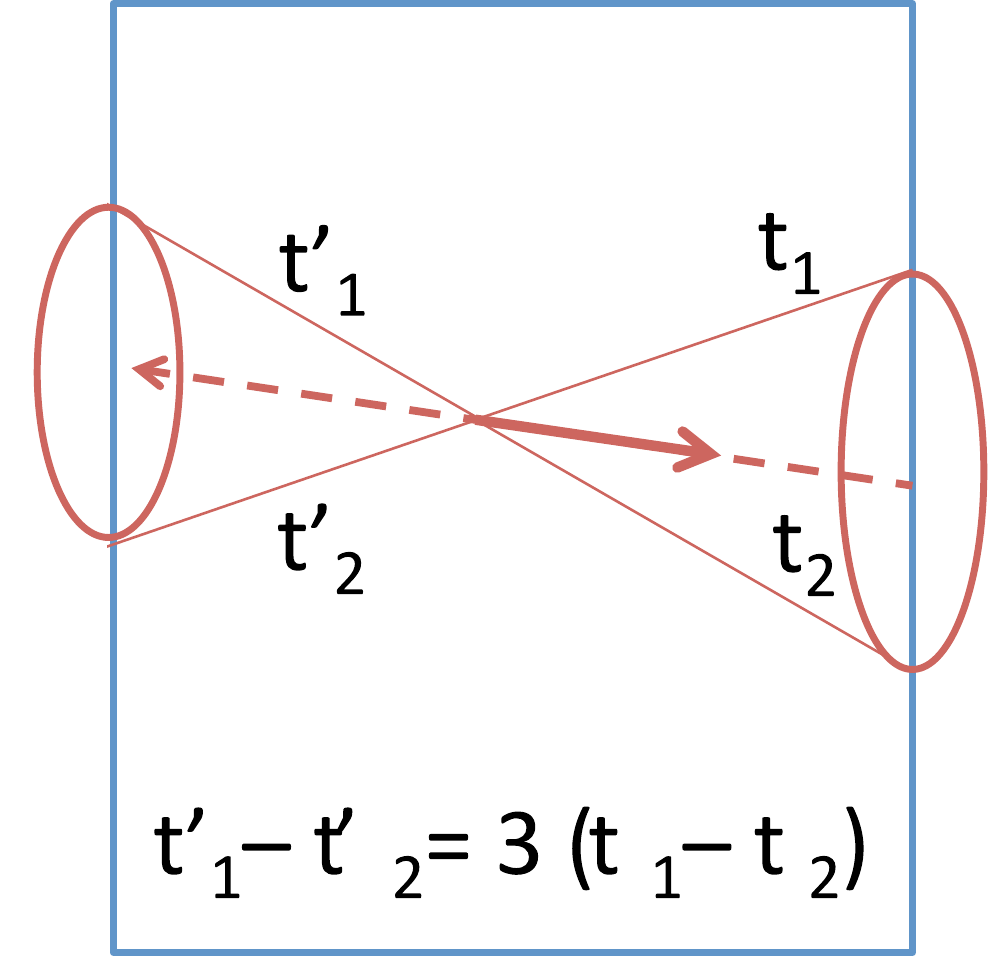}
\end{small}
\caption{a. Parallax gives sensitivity to photon direction. b. Longer photon path length means better sensitivity to timing differences.}
\label{fig:06-origin}
\end{center}
\end{figure}

\begin{em}Sensitivity to photon direction.\end{em} --- Without reflectors, the ring shape is almost degenerate to a simultaneous change of transverse vertex shift ($y$, $z$) and direction change ($\phi$, $\theta$ respectively). While timing information resolves this degeneracy to some extent, with the addition of reflectors the antipodal ring moves over big distances for such a transformation, breaking this degeneracy even stronger (Fig.~\ref{fig:06-origin}a). One can say, the reflectors add sensitivity to the direction of the photons in the primary ring. This mechanism explains the great enhancement of transverse vertex and angle resolution, as well as their similar dependence on the reflectivity.

\begin{em}Sensitivity to timing differences.\end{em} --- Normally vertex sensitivity comes from the timing difference $t_1 - t_2$ of two points on the rings. The reflected light has a much longer path length (Fig.~\ref{fig:06-origin}b) --- about 3x for a particle generated at center of tank --- which means even for the method of vertex reconstruction using timing information, the antipodal ring can contribute to an enhanced resolution.

By varying the timing resolution of the PMTs, we find without reflectors a strong variation of ring reconstruction resolution, while with reflectors the resolution starts to improve only once the timing resolution is improved significantly (about factor 10) beyond values for the SuperK PMTs. Thus we believe the first is a stronger effect.

By being sensitive to photon direction, the antipodal ring might also benefit for particle-identification. For muon-like tracks the photon direction vectors should be more aligned, while for electron-like showers one expects more variance. The enhanced ring reconstruction resolution should also benefit in the particle-identification using kinematic selection of multi-ring events.

\section{Sensitivity at different photo-coverages}

\begin{figure}
\begin{center}
\begin{small}
\includegraphics[width=2.6in]{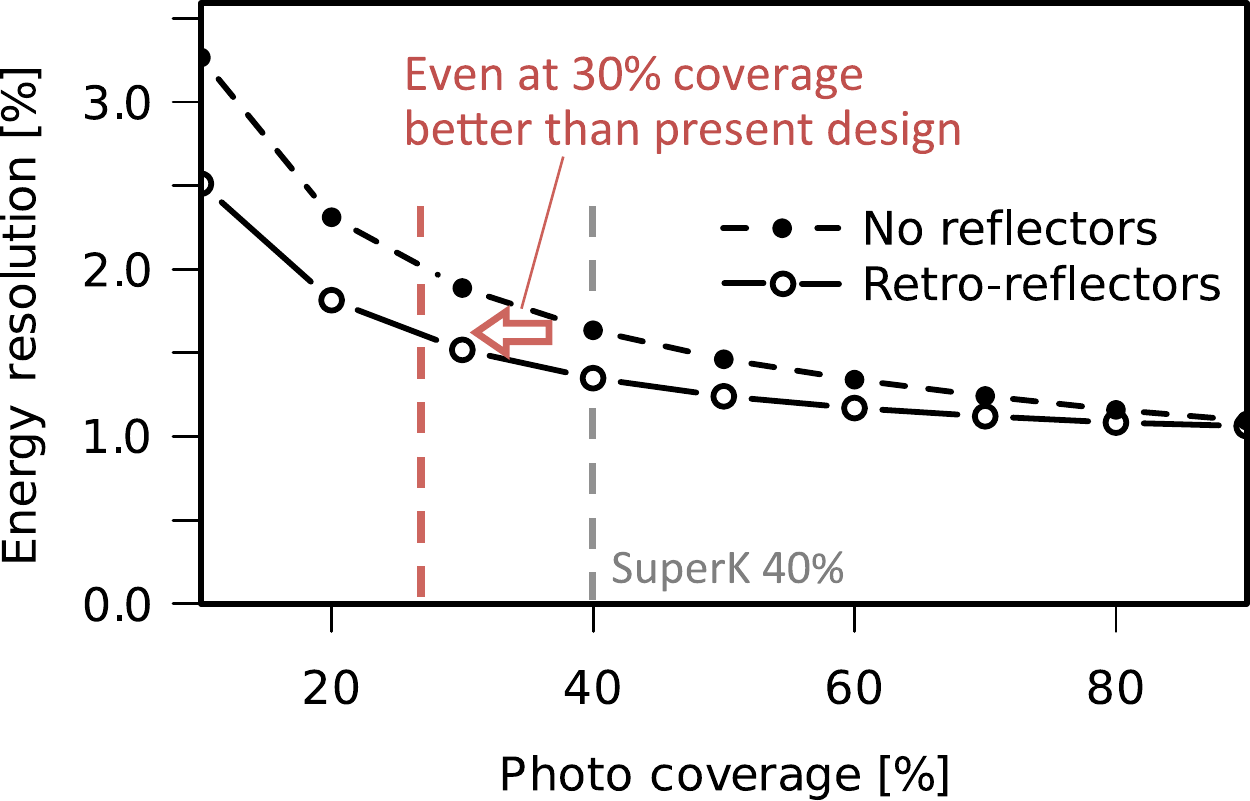}
\includegraphics[width=2.6in]{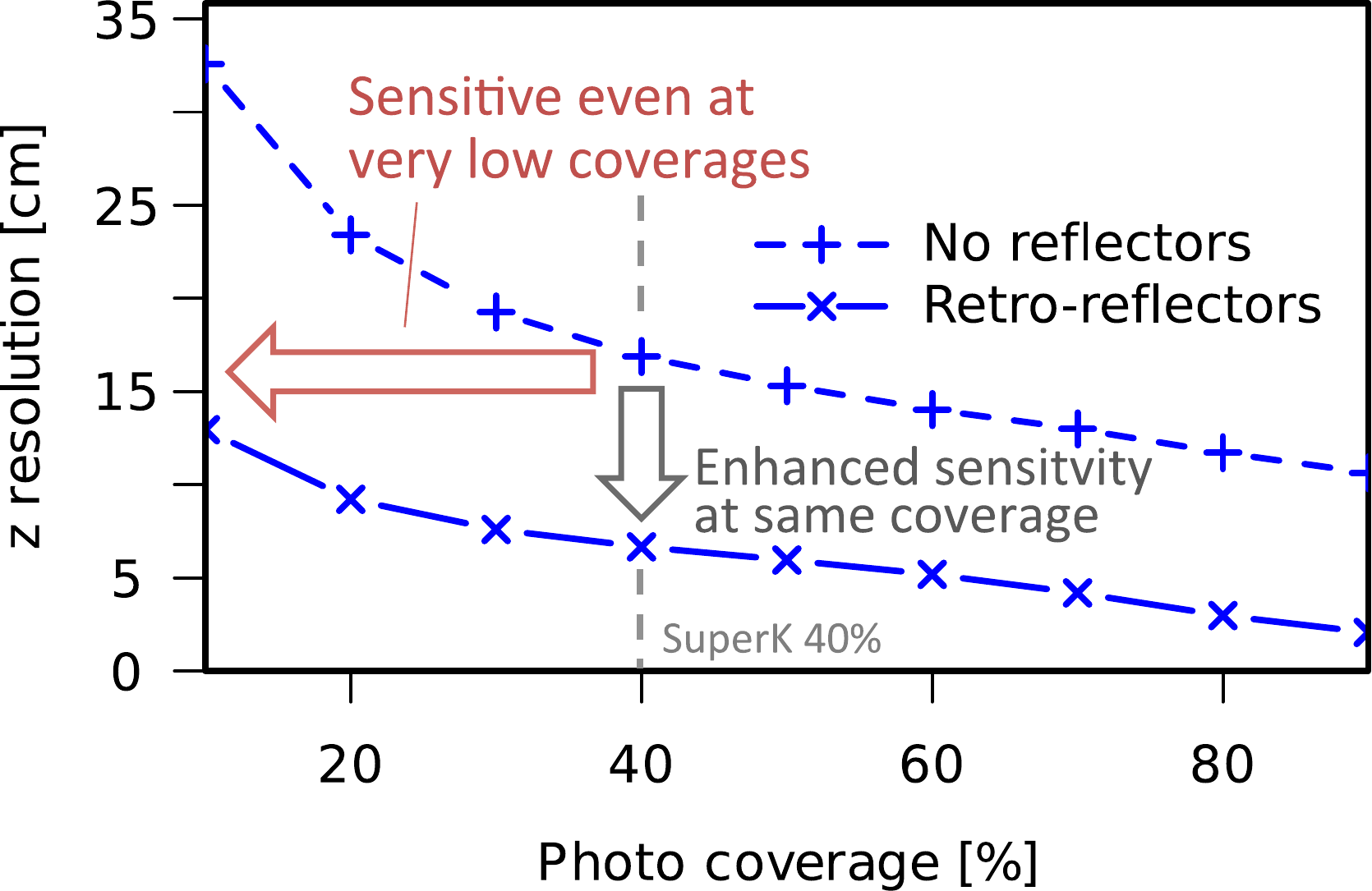}
\end{small}
\caption{Representative ring reconstruction sensitivities ($E$, $z$) for longitudinal ($x$, $E$; left) and transverse ($y$, $z$, $\theta$, $\phi$; right) parameters, with varying photo-coverage.}
\label{fig:07-photo-cov}
\end{center}
\end{figure}

Since the area that can be covered by reflectors increases as we reduce the photo-coverage, the enhancement due to reflectors might be even more significant at smaller photo-coverages. The calculated ring reconstruction resolution for different photo-coverages is shown in Fig.~\ref{fig:07-photo-cov}.

The energy resolution mostly depends on the number of collected photons. The more PMTs we have the less reflectors we can place, so the no-reflector and retro-reflector sensitivities converge for large photo-coverages. While the improvement is not as big as for the transverse observables, one can still reduce the photo-coverage somewhat below 30\% while retaining the same performance as the 40\% coverage of SuperK without reflectors.

The enhanced transverse position and angle sensitivities due to parallax result from a new information class (photon direction), and are thus effective for a wide range of photo-coverages. Even at coverages as low as 10\% the resolution with reflectors is still better than at 40\% without reflectors.

The improved reconstruction resolution means one can reduce the number of PMTs by introducing retro-reflectors, significantly reducing the cost of water-Cherenkov detectors\footnote{Commercially available retro-reflectors with reasonable performance ($<1^\circ$ retro-reflection resolution) cost about 1/10 of SuperK PMTs.}.

\section{Practical challenges}

\begin{figure}
\begin{center}
\begin{small}
\includegraphics[width=2in]{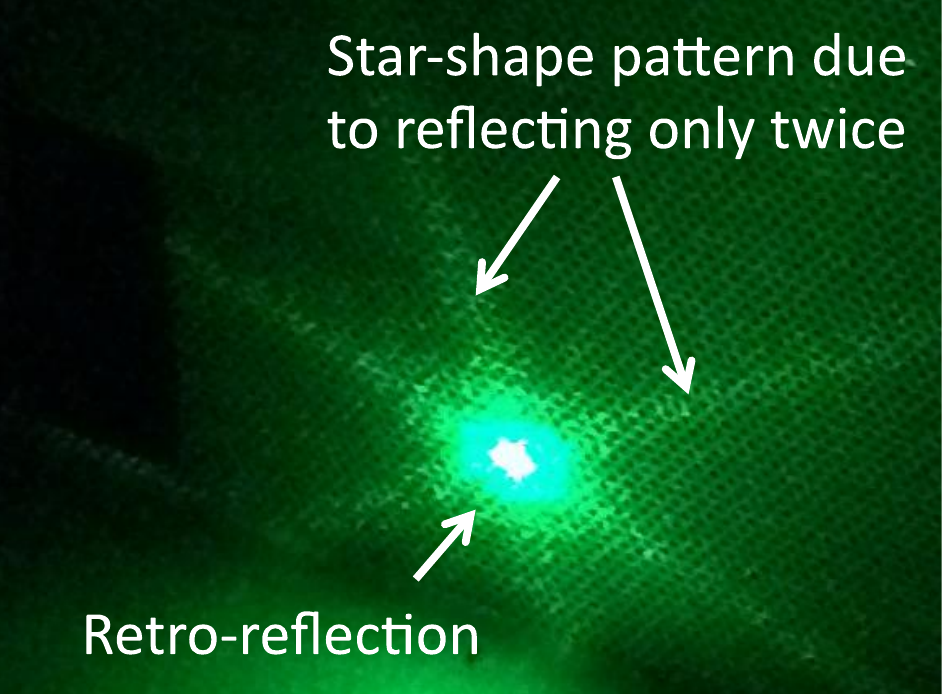}
\end{small}
\caption{Retro-reflection pattern from a laser beam pointed at a commercially available corner-cube micro-prism retro-reflector tape (ORAFOL\textsuperscript{\textregistered} AP1000DL). The bright spot in the center is the retro-reflection. Though not visible here, at the center of the retro-reflection pattern is the pin-hole, through which the laser beam is collimated. The star-shape pattern is due to double-reflection; the ordinary reflection is off-screen and not visible. The dotted patterns are due to the punched aluminum plate that is used both as pin-hole and screen.}
\label{fig:08-patterns}
\end{center}
\end{figure}

Corner cube retro-reflectors provide the most accurate retro-reflection. However, if the incident light reflects less than three times, it will not reflect back into the original direction. For a single reflection and surface reflections for reflective tapes, the reflection is similar to ordinary mirrors. For a double-reflection, it will be retro-reflective along one axis and ordinary along the other. Because corner cube retro-reflectors have a three-fold rotation symmetry, this double-reflection creates a star-shaped pattern (Fig.~\ref{fig:08-patterns}). Such non-retro-reflections reduce the brightness of the retro-reflection, and cause the issues associated with non-retro reflections described in Sec.~\ref{sec:background}. Suppression methods (e.g.~blinds) or calibration methods (e.g.~using the star-shaped reflection pattern) might need to be developed.

\begin{figure}
\begin{center}
\begin{small}
\includegraphics[width=3in]{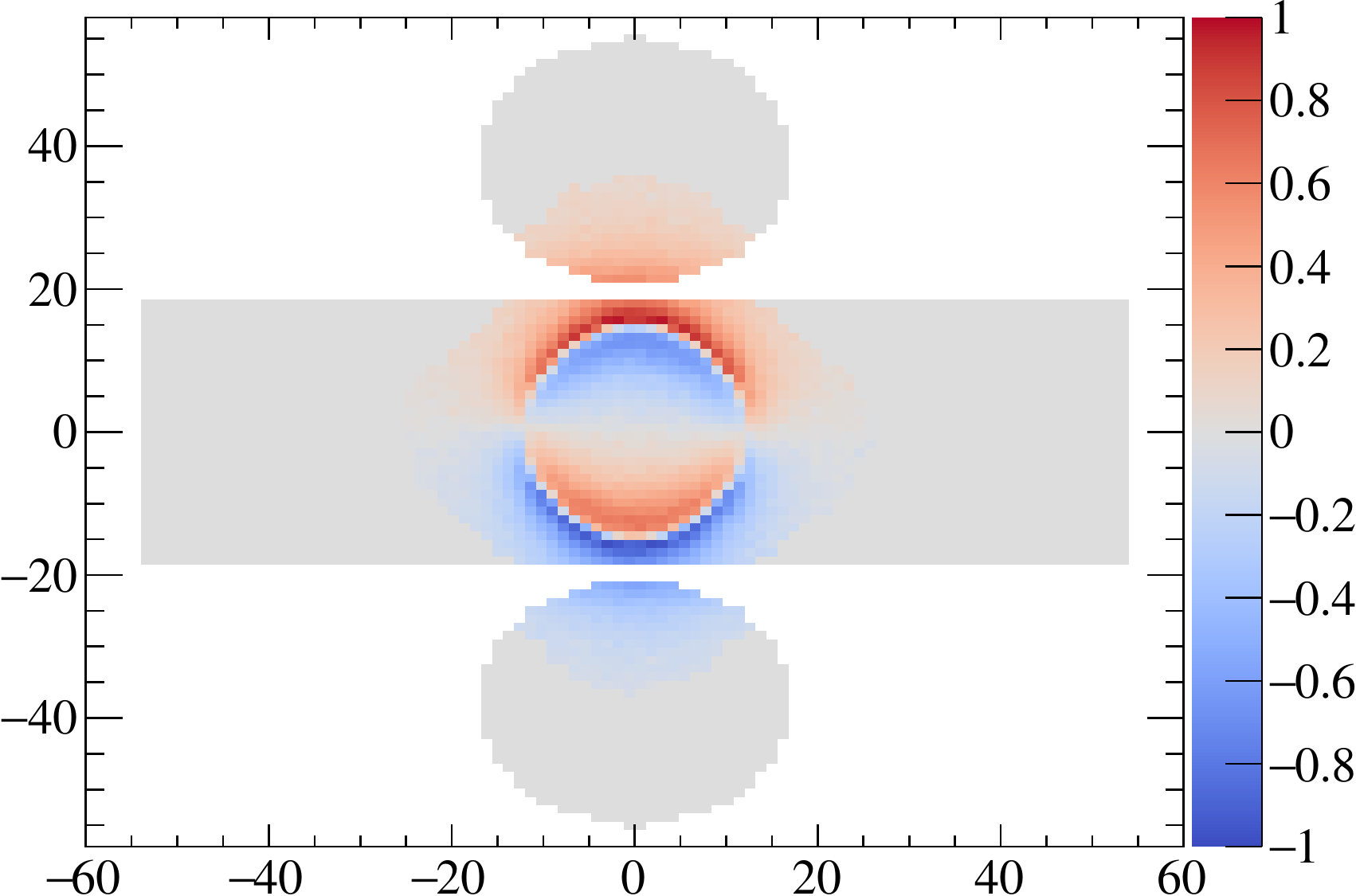}
\includegraphics[width=3in]{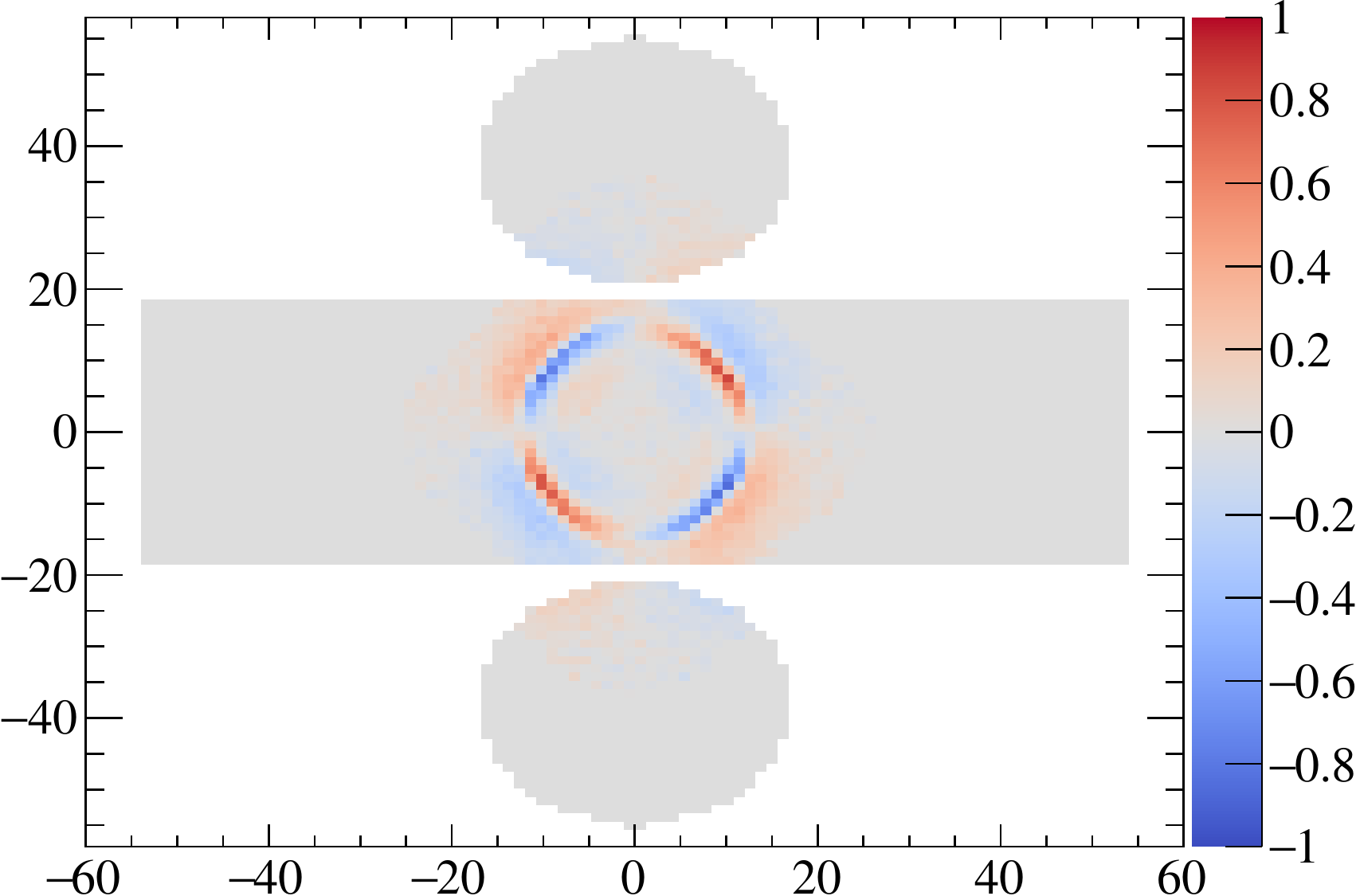}
\end{small}
\caption{Two principal components (normalized eigenvectors of the PMT correlation matrix) for a side-going electron of $500\un{MeV}$, showing dipole-like~(left) and quadrupole-like~(right) deformations of the ring. These are variations existing for identical incident particles. By fitting these modes with large variations, the likelihood fit used for ring reconstruction should improve.}
\label{fig:08-eigenvectors}
\end{center}
\end{figure}

Preliminary ring fitting tests using standard approximations employed in current ring fitting methods did not show improvements to the extent calculated in the likelihood based method above. In particular, it is commonly assumed that for a given particle type (electron, muon, etc.) of certain momentum, the distribution of Cherenkov photons is well described using the average distribution. It seems that for the antipodal ring, the modeling of variations is essential, possibly since the antipodal ring is rather faint while being sensitive to varying photon directions. With a simple modeling of point-to-point variations using principal component analysis (Fig.~\ref{fig:08-eigenvectors}), we see performance improvements similar to the earlier analysis at the expense of higher computational cost. Practically feasible methods for ring-fitting including the effect of Cherenkov profile variations might need to be developed. The modeling of variations could also benefit ring-reconstruction performance for conventional water-Cherenkov detector designs.

\section{Summary}

We proposed a new design for water-Cherenkov detectors with retro-reflectors between PMTs to produce a secondary antipodal ring. Simple simulations suggest an improvement of transverse vertex and angular resolution by up to factor two, mostly due to sensitivity to the direction of the Cherenkov photons from mapping photons originally lost between PMTs onto the other side of the tank. Problems like multiple reflections and alignment difficulties are elegantly solved by using retro-reflectors instead of normal mirrors. For a 40\% photo-coverage detector, this design allows removing a quarter of the PMTs while still providing superior reconstruction performance, reducing the cost of water-Cherenkov detectors significantly. We seek to continue with further studies of particle-identification performance, and development of practical implementations and reconstruction techniques.


\end{document}